\documentclass[english,showpacs,floatfix,nofootinbib,prd]{revtex4}
\usepackage[utf8]{inputenc}
\usepackage[T1]{fontenc}
\usepackage{lmodern}
\usepackage{amsmath}
\usepackage{amssymb}
\usepackage{graphicx}
\usepackage{esint}
\usepackage{longtable}
\usepackage{dcolumn}
\usepackage{babel} 
\usepackage{csquotes} 

\begin{document}

\title{Bouncing cosmology inspired by regular black holes}
\author{J. C. S. Neves}
\email{nevesjcs@ime.unicamp.br}
\affiliation{Instituto de Matemática, Estatística e Computação Científica, Universidade
Estadual de Campinas \\
 CEP. 13083-859, Campinas, SP, Brazil}

\begin{abstract}
In this article, we present a bouncing cosmology inspired by a family of regular black holes. This scale-dependent cosmology deviates from the cosmological principle by means of a scale factor which depends on the time and the radial coordinate as well. The model is isotropic but not perfectly homogeneous. That is, this cosmology describes a universe almost homogeneous only for large scales, such as our observable universe.
\end{abstract}


\maketitle

\section{Introduction}\label{Intro}
Singularities within the General Relativity (GR) context are still an open question in both cosmology \cite{Novello} and black holes (BH) physics \cite{Joshi}. A final answer could be the insertion of quantum effects to solve this problem. However, in several cases, it is possible to avoid the singularities with aid of violations in energy conditions. Such violations are more acceptable since the observation of cosmic acceleration \cite{Supernova,Supernova2}. By assuming these violations, the Hawking-Penrose theorems are not valid. Then, in both cosmology and BH physics, one may construct gravitational field solutions without a singularity at $t=0$, initial time, or $r=0$, the coordinate origin.

The Bardeen \cite{Bardeen2} solution was the first regular black hole (RBH) developed. It is a compact object with an event horizon and without a physical singularity (see, for example, \cite{Ansoldi} and \cite{Lemos_Zanchin} for an introduction). This  achievement is a consequence of ideas from Sakharov and co-authors \cite{Sakharov,Gliner} that the spacetime inside the horizon, where the matter has a high pressure, is de Sitter-like. That is, at the center there is a de-Sitter core which avoids the existence of a singular point. The Bardeen solution has spherical symmetry and it does not violate the weak energy condition (WEC). Later, Hayward \cite{Hayward} has constructed another regular solution with this symmetry. Such as the Bardeen RBH, it does not violate the WEC (recently \cite{Neves}, we present a spherical regular solution which violates the WEC). Currently, regular solutions with axial symmetry (BHs with rotation) have been created as well \cite{Various_axial,Various_axial2,Various_axial3,Various_axial4,Various_axial5,Neves2}. In our work \cite{Neves_Saa}, we have constructed a general class of regular solutions by using a mass function which depends on $r$, the rotation and a cosmological constant.

In another front, physicists and cosmologists have developed several ways to avoid the initial singularity problem, i.e., the big bang. In GR context, in general, bouncing cosmologies violate some energy condition, such as the strong energy condition (SEC). According to these researches, sophisticated bouncing cosmologies\textemdash{}models beyond the standard $\Lambda$CDM model\textemdash{}manage to avoid the big bang problem and the standard model problems (flatness, homogeneity, horizon and isotropy problems), which are solved by adopting the inflationary mechanism by the $\Lambda$CDM model. In this list beyond the standard inflationary model, there exist, for example, the ekpyrotic cosmology \cite{Lehners} and the matter bounce model \cite{Brandenberger}. Moreover, according to Ref. \cite{Steinhardt}, several bouncing cosmologies without the inflationary mechanism do not suffer from the initial conditions and multiverse-unpredictability problems.     

In bouncing cosmology, attempts have been made to construct regular models by ignoring the homogeneity. And this assumption is appropriate from the observational point of view \cite{Wu}. That is, at large scales our universe is described as almost homogeneous, but  the inhomogeneity increases at small scales (see Ref. \cite{Bolejko} for a review in inhomogeneous cosmology). For example, Ruiz and Senovilla \cite{Ruiz_Senovilla} show a class of bouncing or regular cosmologies by means of an inhomogeneous perfect fluid. These results generalize a previous work by Senovilla \cite{Senovilla}. 

Inspired by RBHs, in this work we are mainly interested to study the initial singularity problem by presenting a simple inhomogeneous and scale-dependent bouncing model. This simple model assumes a generalized scale factor similar to the mass function adopted in RBHs (with $r$-dependence) and a barotropic equation of state (EoS). The term in the metric which carries the scale-dependency may determine the regularity of this cosmology. Thus, the model may be regular for either a dust or a radiation dominated universes. However, the model may have a negative energy density at the bounce. According to \cite{Nemiroff}, this feature may be attributed to quantum effects when the universe was very small. In this sense, the energy conditions are not valid.  

The structure of this paper is as follows: in Section II we present some features of RBHs with spherical symmetry; in Section III we show the bouncing model with a $t$ and $r$-dependence in the scale factor; in Section IV the final remarks are presented. In this work, we have used the metric signature $diag(-+++)$ and the geometric units $G=c=1$, where $G$ is the gravitational constant and $c$ is the speed of light in vacuum. 

\section{Some features of spherical RBHs}\label{Feat}
To construct general RBHs, we adopted in Ref. \cite{Neves_Saa} a general mass function
\begin{equation}
m(r)=\frac{M_{0}}{\left[1+\left(\frac{r_{0}}{r}\right)^{q}\right]^{p/q}},\label{Mass_term}
\end{equation}
where $M_{0}$ and $r_{0}$ are, respectively, mass and length parameters. To accommodate event horizon(s), one assumes $r_0 \ll M_0$ in Eq. (\ref{Mass_term}). The Bardeen and Hayward RBHs correspond to $p=3,q=2$ and $p=3,q=3,$
respectively. Our solution generalizes previous works because it adopts the cosmological constant, axial symmetry and the general mass term (\ref{Mass_term}). With this general mass function it is possible to obtain RBHs with either spherical symmetry or axial symmetry. Moreover, we show that the WEC is always violated when the rotation (axial symmetry )is present.

The only one possible value of $p$ in Eq. (\ref{Mass_term}) is 3, because the metrics with the mass
function (\ref{Mass_term}), in the limit of small $r$, are de Sitter-like with this value for $p$, i.e., these geometries have a de-Sitter core at the central region only for that choice. 

With  aid of the mass function (\ref{Mass_term}) and the general metric
\begin{equation}
ds^{2}=-A(r)dt^{2}+\frac{1}{B(r)}dr^{2}+r^{2}d\Omega^{2},\label{metric}
\end{equation}
where $d\Omega^{2}=d\theta^{2}+\sin^{2}\theta d\phi^{2},$ an entire family of regular spherical solutions may be obtained (in the axial case, the approach is different, and it is necessary, for example, according to Ref. \cite{Neves_Saa}, the Kerr-Schild ansatz). These metrics are not vacuum solutions of the gravitational field equations. By using the so-called Synge's $g$-method (applied, for example, in Refs. \cite{Neves_Saa,Hernandez}), the energy-momentum tensor, when $A(r)=B(r)=1-2m(r)/r$, is given from the metric (\ref{metric}):
\begin{equation}
T_{\nu}^{\mu}=\frac{1}{8\pi}\left(\begin{array}{cccc}
-\frac{2m'(r)}{r^{2}}\\
 & -\frac{2m'(r)}{r^{2}}\\
 &  & -\frac{m''(r)}{r}\\
 &  &  & -\frac{m''(r)}{r}
\end{array}\right),\label{Energy-momentum}
\end{equation}
where the symbol (') represents an ordinary derivative with respect to $r$. With the energy-momentum tensor in the form $T_{\nu}^{\mu}=diag\left(-\rho,P_{r},P_{t,}P_{t}\right)$, where $\rho$, $P_{r}$ and $P_{t}$ are the energy density, radial pressure and tangential pressure, respectively, these metrics are not isotropic ($P_{r}=P_{t}$) for a general $m(r)$.

The standard RBHs (Bardeen and Hayward solutions) do not violate the WEC. That is, one has  $\rho\geq0$ and $\rho+P_{i}\geq0$, for $i=r$ and $i=t$, in these metrics. In these solutions, $A(r)=B(r)$ leads to $\rho+P_{r}=0$, $\rho=m'(r)/(4\pi r^2)$ and $\rho+P_{t}=(2m'(r)-rm''(r))/(8\pi r^{2})$, which are positive definite.

The most important feature in the standard Bardeen and Hayward metrics is the regularity, i.e., the scalars are finite everywhere. For example, for $r\rightarrow 0$, the Bardeen and Hayward black holes have the Kretschmann scalar written as 
\begin{equation}
\lim_{r\rightarrow 0} K(r) = \lim_{r\rightarrow 0} R_{\mu\nu\alpha\beta}R^{\mu\nu\alpha\beta}=96\left(\frac{M_{0}}{r_{0}^3}\right)^2,\label{K_Bardeen}
\end{equation}
by using the notation of Eq. (\ref{Mass_term}), where this limit is valid for any $q$ positive integer.    

\section{Bouncing cosmology inspired by RBHs}
In GR context, bouncing models may be obtained by using the Friedmann-Lemaître-Robertson-Walker (FLRW) metric in the $(t,r,\theta,\phi)$ coordinates. In our work, one has a flat FLRW-type cosmology, 
\begin{equation}
ds^2=-dt^2+A(t,r)^2 \left(dr^2+r^2 d\Omega^2 \right). 
\label{FLRW}
\end{equation}
The dimensionless scale factor $A(t,r)$ of this model depends on $r$ as well. In this sense, the cosmological principle is not valid. We assume a function $A(t,r)$ similar to the mass function used to obtain regular black holes. That is,
\begin{equation}
A(t,r) = \frac{a(t)}{\left[1+\left( \frac{b(t)}{r} \right)^q \right]^{p/q}},
\label{A(t,r)}
\end{equation}
where, as we shall see, the dimensionless function  $a(t)$ is directly related to the spacetime dynamics, and the function $b(t)$ is essential to obtain a bounce. If we interpret $a(t)$ and $b(t)$ as constants in $r$, the similarity between Eq. (\ref{Mass_term}) and Eq. (\ref{A(t,r)}) will be adequate. 

By using the metric (\ref{FLRW}) in the Einstein equations $G^{\mu}_{\nu}=8\pi T^{\mu}_{\nu}$,  the components of the Einstein tensor are written as:
\begin{align}
G^{0}_{0}& = - 3\left(\frac{\dot{A}}{A}\right)^2-\left(\frac{A'}{A^2}\right)^2+2\left(\frac{A''}{A^3}\right)+4\left(\frac{A'}{rA^3}\right), 
\label{G00}\\
G^{0}_{1}& = 2\left[\frac{\dot{A'}}{A} -\frac{\dot{A}A'}{A^2}\right], 
\label{G01} \\
G^{1}_{0}& = -2\left[\frac{\dot{A'}}{A^3}-\frac{\dot{A}A'}{A^4}\right],
\label{G10} \\
G^{1}_{1}& = -2\left(\frac{\ddot{A}}{A}\right)-\left(\frac{\dot{A}}{A}\right)^2+\left(\frac{A'}{A^2}\right)^2+2\left(\frac{A'}{rA^3}\right), 
\label{G11}\\
G^{2}_{2}&=G^{3}_{3}= -2\left(\frac{\ddot{A}}{A}\right)-\left(\frac{\dot{A}}{A}\right)^2-\left(\frac{A'}{A^2}\right)^2+\left(\frac{A''}{A^3}\right) +\left(\frac{A'}{rA^3}\right),
\label{G22}
\end{align}
where dot stands for derivative w.r.t. time $t$. We are looking for a cosmic fluid with two features: isotropy and inhomogeneity. The first feature determines the values of $p$ and $q$ in Eq. (\ref{A(t,r)}) because isotropy states that $T^{1}_{1}=T^{2}_{2}=T^{3}_{3}$, i.e.,
\begin{equation}
G^{1}_{1}-G^{2}_{2}=0,
\end{equation}
that implies to say that 
\begin{align}
\left[2+q+(2+p)\left(\frac{b(t)}{r}\right)^q\right]=0.
 \label{Isotropy}
\end{align}
Then, independently of the function  $b(t)$, the possible values to generate an isotropic cosmic fluid are $p = q = -2$ in Eq. (\ref{A(t,r)}). Besides, with these values for $p$ and $q$, the FLRW metric is restored in the limit $b(t) \rightarrow \infty$.
 
The form of the function $A(t,r)$  implies off-diagonal energy-momentum tensor components ($T_{10}=T_{01}\neq 0$), which indicate the momentum density. This form forbids a factorable $A(t,r)=a(t)F(r)$, which in some sense is a FLRW metric by using a specific coordinate transformation. A scale factor, within an inhomogeneous cosmology, with these features was studied in Ref. \cite{Kim}. However, the authors did not consider a bouncing model. 

From the discussion above, we basically seek an energy-momentum tensor given by
\begin{equation}
T^{\mu}_{\nu}=\left(\begin{array}{cccc}
-\rho & T_{1}^{0}\\
T_{0}^{1} & P\\
 &  & P\\
 &  &  & P
\end{array}\right).
\label{Energy-momentum2}
\end{equation}
As the energy-momentum tensor given by Eq. (\ref{Energy-momentum2}) is not diagonal, it does not describe a perfect fluid. According to \cite{Kim}, a fluid of this type may be described by a viscous fluid. However, as we shall see, with a suitable choice for $b(t)$, it is possible to obtain a diagonal metric away from $t=0$, i.e., the terms $T_{01}$ and $T_{10}$ may be relevant only near the bounce. Moreover, with an appropriate choice of $b(t)$, the cosmic fluid written as (\ref{Energy-momentum2}) renders both regular energy density and pressure at $t=0$.

\subsection{Conditions for a bounce}\label{Conditions}
In general, the existence of a bounce depends on the conditions for the scale factor. To have a minimum value at $t=0$, with the metric given by Eq. (\ref{FLRW}), the function $A(t,r)$ must obey the following criteria at the bounce:
\begin{itemize}
\item[(i)]$A(t=0,r)=A_{min}>0$, 
\item[(ii)]$\dot{A}(t=0,r)=0$,
\item[(iii)]$\ddot{A}(t=0,r) > 0$.
\end{itemize}  
The condition (i) indicates a minimum value for the scale factor and leads to a minimum value $a_{min}>0$. The condition (ii) implies $\dot{a}(0)=\dot{b}(0)=0$. The third leads to 
\begin{equation}
\frac{\ddot{b}(0)}{b(0)^3}>-\frac{\ddot{a}(0)}{2r^2 A_{min}},
\end{equation} 
and $\ddot{a}(0)>0$. These conditions give some clues for finding an appropriate $b(t)$.  

\subsection{Energy-momentum tensor components} \label{EMT_components}
The energy-momentum tensor given by Eq. (\ref{Energy-momentum2}) satisfies the energy conservation, i.e., 
\begin{equation}
\nabla_{\nu} T^{\mu\nu}=0.
\end{equation}
And the equation above, by using the barotropic EoS, 
\begin{equation}
P=w\rho,
 \label{EoS}
\end{equation}
renders a relation between the energy density and the scale factor. That is,
\begin{equation}
\rho(t,r) \propto A(t,r)^{-3(1+w)}.
\label{rho_a}
\end{equation}
As we can see, the relation (\ref{rho_a}) is similar to the relation between the energy density and scale factor in the FLRW metric. For a radiation dominated universe, $w=1/3$ implies $\rho \propto A^{-4}$. Then, after the bounce, when the scale factor is small, the radiation dominates the cosmic expansion over, for example, the matter component, which by using $w=0 $ gives $A^{-3}$ in Eq. (\ref{rho_a}).     

We propose a departure from the cosmological principle: in particular, a deviation from the homogeneity. At a time $t$, the homogeneity states the following:
\begin{equation}
\rho (r)=\rho (r+r_c)=\rho,  
\end{equation}   
for a comoving distance $r_c$ which separates two observers. Therefore the energy density is constant for all observes. With the scale factor given by Eq. (\ref{A(t,r)}), we assume that
\begin{equation}
\rho (r_l) \approx \rho,
\end{equation} 
where $r_l$ stands for large values of $r$. Thus the universe is almost homogeneous for large scales. The value of $r$ is more determinant on $\rho$ when the scale is smaller.

The metric (\ref{FLRW}) with aid of the Einstein equations (or specifically the Eq. (\ref{G00})) renders the expression for $\rho (t,r)$, which is considered one of the Friedmann-type equations: 
\begin{equation}
8\pi\rho = 3\left(\frac{\dot{a}}{a}\right)^2+\frac{12}{(ab)^2}+\frac{12 \dot{a}\dot{b}}{ab\left[1+\left(\frac{b}{r}\right)^2 \right]}+\frac{12\dot{b}^2}{b^2\left[1+\left(\frac{b}{r}\right)^2 \right]^2}.
\label{rho}
\end{equation}
By using the expansions for large, $r_l$, and small, $r_s$, values of the radial coordinate, the terms in Eq. (\ref{rho}) which depend on the radial coordinate are written as
\begin{align}
\frac{1}{1+\left(\frac{b}{r_s}\right)^2}& \approx \left(\frac{r_s}{b} \right)^2,
\label{Small}
 \\
\frac{1}{1+\left(\frac{b}{r_l}\right)^2}& \approx 1-\left(\frac{b}{r_l} \right)^2.
\label{large}
\end{align}
The variations of Eqs. (\ref{Small})-(\ref{large}) are $\frac{d}{dr_s}(\frac{r_s}{b})^2=\frac{2r_s}{b^2}$  and $\frac{d}{dr_l}(1-(\frac{b}{r_l})^2)=\frac{2b^2}{r^{3}_{l}} $, respectively. Thus, it is worth noting that for large $r$ the variation is small. Therefore the metric (\ref{FLRW}) is able to describe a universe which is homogeneous or almost $r$-independent at large scales. 

In the same way, by using the Einstein tensor components (\ref{G11}) or (\ref{G22}), one has the pressure as a function of $t$ and $r$:
\begin{align}
8\pi P & = -\left(\frac{\dot{a}}{a}\right)^2-2\left(\frac{\ddot{a}}{a}\right)-\frac{4}{(ab)^2}- \frac{4\left(3\dot{a}\dot{b}+a \ddot{b}\right)}{ab\left[1+\left(\frac{b}{r}\right)^2 \right]}-\frac{4\dot{b}^2\left[2- \left(\frac{3b}{r}\right)^2\right]}{b^2\left[1+\left(\frac{b}{r}\right)^2 \right]^2}.
\label{P}
\end{align}

It is interesting to note that the off-diagonal elements in the energy-momentum tensor are almost null for both small and large scales. These elements are written as
\begin{equation}
T_{01} =  T_{10}=-\frac{b\dot{b}}{\pi r^3\left[1+\left(\frac{b}{r}\right)^2 \right]^2},
\end{equation}
and its limits, for small and large scales, are
\begin{equation}
\lim_{r \rightarrow 0} T_{01}= \lim_{r\rightarrow \infty} T_{01}=0.
\label{Limits_off_diagonals}
\end{equation}
Moreover, it is possible to build a function $b(t)$ that makes these components non-vanishing only at the bounce, $t=0$, for example, $b(t)\propto e^{t^2}$. Therefore, today, the cosmic fluid is considered almost a perfect fluid, and the matrix (\ref{Energy-momentum2}) is pretty much diagonal.   

\subsection{The function $b(t)$}\label{Function_b}
In terms of the regularity of $ds^2$, $A(t,r)$, $\rho$ and $P$, it is worth noting that the function $b(t)$  plays an important role. According to the above discussion on the homogeneity, the function $b(t)$ must increase with the time. Because the homogeneity in the past was possible at smaller scales than today (the universe was more homogeneous in the past). 

By adopting a specific function $b(t)$, we may investigate the real meaning for the function $A(t,r)$ or $a(t)$. The line element (\ref{FLRW}) provides the following interpretation: at a fixed time $t$ with $\theta$ and $\phi$ constants, the proper distance between two observers, the first at the origin and the second at a comoving distance $r_c$, reads
\begin{align}
d(t,r)=& \ a(t)\int^{r_c}_{0}\frac{dr}{\left[1+\left(\frac{r}{b(t)}\right)^2 \right]} \nonumber \\ 
            =& \ a(t)b(t) \arctan \frac{r_c}{b(t)} .
\label{ds}            
\end{align}
For  functions $b(t)$ and $a(t)$ which increase or decrease with the time, the proper distance between the two observes increases or decreases, respectively, with the time, i.e., the function $a(t)$ in our metric plays the same role to determine proper distances such as in the FLRW metric. We shall see that, contrary to the standard cosmological model, in our model the proper distance $d(t,r)$ does not diverge at $t=0$ because both $a(t)$ and $A(t,r)$ are regular.

\subsection{Scale factor}\label{Scale_factor}
The second Friedmann-type equation comes from $G^{i}_{i}$ (with $i$ assuming the values 1, 2 or 3 due to the isotropy) and the Eq. (\ref{rho}). Then, one has
\begin{equation}
\frac{\ddot{a}}{a}=-\frac{4 \pi}{3}\left(\rho+3P \right)-\frac{2\left( 2\dot{a}\dot{b}+a\ddot{b}\right)}{ab \left[1+\left(\frac{b}{r}\right)^2 \right]}-\frac{2\dot{b}^2 \left[1-3 \left(\frac{b}{r}\right)^2 \right]}{b^2 \left[1+\left(\frac{b}{r}\right)^2 \right]^2}.
\label{Friedmann2}
\end{equation}
The conditions at the bounce, i.e., $\dot{a}(0)=\dot{b}(0)=0$ and $\ddot{a}(0)>0$, lead to the following inequality with aid of Eq. (\ref{Friedmann2}) and the EoS (\ref{EoS}):
\begin{equation}
\frac{2\pi}{3}\left(\rho+3P \right)\bigg|_{t=0} = \frac{\left(1+3w \right)}{\big(a_{min} b(0)\big)^2} < -\frac{\ddot{b}(0)}{b(0)\left[1+\left(\frac{b(0)}{r} \right)^2 \right]}.
\label{Inequality}
\end{equation}
In terms of $b(t)$, for a matter or radiation dominated universe, the inequality above may be satisfied, for example, with $\ddot{b}(0)/b(0) < 0$ or a function written as $b(0)=if(0)$. The latter choice provides the same signal for the term $r^2$ in the denominator in Eq. (\ref{A(t,r)}) such as found in other inhomogeneous metrics, such as Ref. \cite{Kim}. As we shall see, both choices have consequences for the SEC according to the left side of Eq. (\ref{Inequality}).

It is rather important to note the first Friedmann-type equation for small scales. For these values, $r \approx 0$, the Eq. (\ref{rho}) is similar to the standard FLRW metric with a term which is proportional to the inverse of a variable curvature-term, $k(t)$. That is,
\begin{equation}
\left(\frac{\dot{a}}{a} \right)^2 = \frac{8 \pi \rho}{3}-\frac{\frac{4}{b^2}}{a^2}.
\label{Friedmann_small_r}
\end{equation} 
With a large value for $b(t)$ (or $1/b(t) \approx 0$) today, one has a almost flat universe in these scales. This detail confirms our understanding, discussed in subsection \ref{EMT_components}, that the function $b(t)$ increases with the time to generate almost homogeneity at large scales.

Let us solve the equation for $a(t)$ analytically for $r \approx 0$. By using the approximation for small $r$, Eq. (\ref{Friedmann_small_r}), it is possible to obtain an analytical solution for $a(t)$ and, consequently, for $A(t,r)$. Its solution for a radiation dominated universe ($P=\frac{1}{3}\rho$), for example, reads
\begin{equation}
a(t) = \sqrt{C_1 + C_2 t -\int dt \int \frac{8dt}{b(t)^2}}.
\end{equation}
For a FLRW universe, $1/b(t) \approx 0$, one has $a(t)\propto t^{1/2}$.

However, in Fig. 1, one has the shape of $A(t,r)$ by using Eqs. (\ref{rho}) and (\ref{Friedmann2}), which were solved numerically for the barotropic EoS $P=w\rho$. For small scales, our neighbourhood, we can see that the expansion depends strongly on the scale, i.e., for two scales $r_1>r_2$, the expansion is larger for $r_1$. 

\begin{figure}
\begin{center}
\includegraphics[scale=0.52]{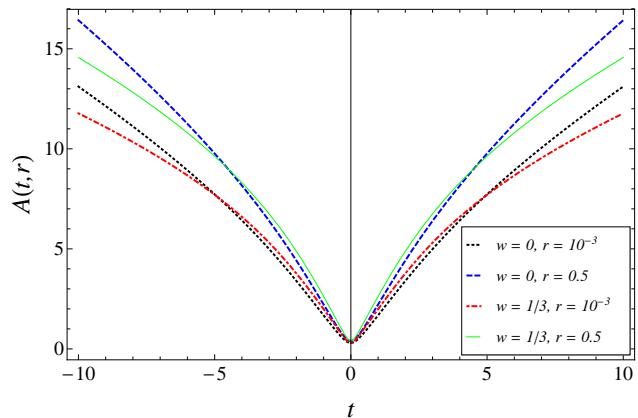}
\end{center}
\caption{The scale factor for matter and radiation dominated universes, $w=0$ and $w=1/3$, respectively, with scales $r=10^{-3}$ and $r=0.5$. The function $b(t)$ used was $ic_{1}e^{c_{2}t^2}$, with $c_1 = 1$ playing the role of a length constant, and $c_2 = 0.01$. As we can see, given two scales $r_1>r_2$ for the same $w$, the expansion is faster for $r_1$. Moreover, by comparing at same scale, for example $r=0.5$, the variation of $A(t,r)$ for small values of $t$ is larger for the radiation dominated universe. This statement agrees with the discussion on the Eq. (\ref{rho_a}).}
\end{figure}

\subsection{Strong energy condition}\label{Energy_Condition}
As we said in Introduction, bouncing cosmologies violate energy conditions in GR context. In general, the SEC is violated. This energy condition claims that $\rho + 3P \geq 0$. It is clear from the inequality (\ref{Inequality}) that to produce a bounce the SEC must be violated if $b(0)=if(0)$, i.e.,
\begin{equation}
\rho+3P<0
\end{equation}
with this choice. 

After the evaluation of the function $A(t,r)$ or $a(t)$, the expressions for the energy density and pressure may be obtained. With a regular $a(t)$, both the energy density and pressure are regular at the bounce. Moreover, the density may be negative near the bounce. This is clear from the expression (\ref{rho}) with the assumptions that $\dot{a}(0)=\dot{b}(0)=0$ and $b(0) = i f(0)$. Then,
\begin{equation}
\rho (0,r) = - \frac{3}{2\pi \big(a_{min} f(0) \big)^2}.
\end{equation} 
One has a result which is $r$-independent. In this sense, within the imaginary choice for $b(t)$, the cosmic fluid given by Eq. (\ref{Energy-momentum2}) does not satisfy the energy condition hypothesis which assumes the positivity of the energy density.   

On the other hand, according to the right side of Eq. (\ref{Inequality}), the condition $\ddot{b}(0)/b(0) < 0$ with a real function $b(t)$ may provide a regular solution which does not violated the SEC.

\subsection{Scalars}
A cosmological bouncing model means a nonsingular cosmology with a minimum scale factor at the bounce \cite{Novello}. In FLRW or FLRW-type metrics, the scale factor determines the regularity of physical quantities such as energy density and pressure. Moreover, for these metrics, geometrical quantities like the Ricci and Kretschmann scalars depend on the scale factor as well. As we can see in the cosmological model presented in this paper, $A(t,r)$ may attain a minimum value, i.e., in our model $A_{min}(0,r)\neq 0$. Thus, according to Eqs.(\ref{rho}) and (\ref{P}), the energy density $\rho$ and the pressure $P$ may be regular at the bounce. Besides, such as in black holes physics, the regularity of geometrical scalars may indicate the spacetime regularity in cosmology as well\footnote{Besides the scalars, another criterion to determine the spacetime regularity depends on the nonexistence of incomplete geodesics. In this sense, a nonsingular or regular spacetime is defined as geodesically complete. As we can see in Wald's book (\cite{Wald}, chapter 9), there are geometries with scalars which diverge but such spacetimes are geodesically complete. An example for this case is found in Ref. \cite{Olmo}, where a wormhole geometry has problem with scalars but it is geodesically complete. However, the standard FLRW cosmology presents the two problems: the geodesics are incomplete and the scalars diverge.}. With the evaluated scale factor or the function $a(t)$,  the Ricci and Kretschmann ($K=R_{\alpha\beta\gamma\delta}R^{\alpha\beta\gamma\delta}$) scalars are obtained for any scale at the bounce:
\begin{align}
R\bigg|_{t=0} & = 6\left[\frac{\ddot{a}}{a_{min}}+\frac{2\left(a_{min}^2b\ddot{b}+\frac{2b}{r^2}+2\right)}{\big(a_{min}b\big)^2\left[1+\left(\frac{b}{r}\right)^2 \right]} \right],
\label{R} \\
K\bigg|_{t=0} & = 12\Bigg[\left(\frac{\ddot{a}}{a_{min}}\right)^2+\frac{4\ddot{a}\ddot{b}}{a_{min}b\left[1+\left(\frac{b}{r}\right)^2 \right]}+\frac{4\left(a_{min}^4 b^2\ddot{b}^2+\frac{4b^4}{r^4}+\frac{8b^2}{r^2}+4 \right)}{\big(a_{min}b \big)^4\left[1+\left(\frac{b}{r}\right)^2 \right]^2} \Bigg].
\label{K}
\end{align}
Both scalars may be regular at the bounce because $a(0)=a_{min}$ and $b(0)$ may be finite and non-null. However, for the case where the function $b(t)$ is imaginary at $t=0$, $b(0)=if(0)$, there exists a singular surface, according to the scale, at $r=f(0)$.  

\section{Final remarks}
RBHs and bouncing cosmologies are two interesting areas in theoretical physics today. By using GR and theories beyond the standard physics, researches have constructed geometries which avoid the problem of the singularities. In this article, we join these two areas. Inspired by an approach in BH physics to generate metrics without singularity, we have investigated a cosmology with a scale factor which depends on $t$ and the scale $r$ as well. The expression for the scale factor resembles to the mass function developed in \cite{Neves_Saa} to obtain RBHs. With an appropriate choice for the free function in the metric, $b(t)$, we have shown a bouncing model which deviates from the cosmological principle because it is described by an inhomogeneous and isotropic cosmic fluid. The model is scale-dependent, and it is almost homogeneous only for large scales, such as the observable universe. At small scales, the interpretation of the Friedmann-type equations is quite simple: a universe with a variable term which is inversely proportional to the curvature. Within these scales, if $r_1>r_2$, the variation of the scale factor is larger for $r_1$.

Lastly, the model must violate the SEC for functions $b(t)$ which are imaginary at the bounce. However, with this choice there exists a singular surface for a specific scale, typically $r=\vert b(0) \vert$, a singularity in the radial coordinate. The same \enquote{horizon} was found in Ref. \cite{Kim} as well.

\begin{acknowledgments}
This work was supported by Fundação de Amparo à Pesquisa do Estado de São Paulo (FAPESP), Brazil (Grant No. 2013/03798-3). I would like to thank Alberto Saa and an anonymous referee for comments and suggestions.
\end{acknowledgments}


\begin{thebibliography}{References}

\bibitem{Novello}Novello, M., Perez Bergliaffa, S. E.: Bouncing cosmologies. Phys. Rept. \textbf{463}, 127 (2008)

\bibitem{Joshi}Joshi, P. S.: Spacetime singularities. In: Ashtekar, A., Petkov, V. (eds) Springer Handbook of Spacetime, p. 409. Springer-Verlag, Berlin Heidelberg (2014)

\bibitem{Supernova}Riess, A. G., et al. (Supernova Search Team Collaboration): Observational evidence from supernovae for an accelerating universe and a cosmological constant. Astron. J. \textbf{116}, 1009 (1998) 

\bibitem{Supernova2}Perlmutter, S.,  et al. (Supernova Cosmology Project Collaboration): Measurements of omega and lambda from 42 high-redshift supernovae. Astrophys. J. \textbf{517}, 565 (1999)

\bibitem{Bardeen2}Bardeen, J. M.: Non-singular general-relativistic collapse. In: Conference Proceedings of GR5, p. 174. Tbilisi, URSS (1968)

\bibitem{Ansoldi}Ansoldi, S.: Spherical black holes with regular center: a review of existing models including a recent realization with Gaussian sources. In: Proceedings of  BH2, Dynamics and Thermodynamics of Black Holes and Naked Singularities. Milano, Italy (2007)

\bibitem{Lemos_Zanchin}Lemos J. P. S., Zanchin, V. T.: Regular black holes: Electrically charged solutions, Reissner-Nordström outside a de Sitter core. Phys. Rev. D \textbf{83}, 124005 (2011)

\bibitem{Sakharov}Sakharov, A. D.: Sov. Phys. JETP \textbf{22}, 241 (1966)

\bibitem{Gliner}Gliner, E. B.: Sov. Phys. JETP \textbf{22}, 378 (1966)  

\bibitem{Hayward}Hayward, S. A.: Formation and evaporation of non-singular black holes. Phys. Rev. Lett. \textbf{96},  031103 (2006)

\bibitem{Neves}Neves, J. C. S.: Deforming regular black holes. Int. J. Mod. Phys. A \textbf{32}, 1750112 (2017) 

\bibitem{Various_axial}Smailagic, A., Spallucci, E.: "Kerrr" black hole: the lord of the string. Phys. Lett. B \textbf{688}, 82 (2010) 

\bibitem{Various_axial2}Modesto, L., Nicolini, P.: Charged rotating noncommutative black holes. Phys. Rev. D \textbf{82}, 104035 (2010) 
 
\bibitem{Various_axial3}Bambi, C., Modesto, L.: Rotating regular black holes. Phys. Lett. B \textbf{721}, 329 (2013)
 
\bibitem{Various_axial4}Toshmatov, B., Ahmedov, B., Abdujabbarov, A., Stuchlik, Z.: Rotating regular black hole solution. Phys. Rev. D \textbf{89}, 104017 (2014) 
 
 \bibitem{Various_axial5}Azreg-Ainou, M.: Generating rotating regular black hole solutions without complexification. Phys. Rev. D \textbf{90}, 064041 (2014)
 
\bibitem{Neves2}Neves, J. C. S.: Note on regular black holes in a brane world. Phys. Rev. D \textbf{92}, 084015 (2015)  

\bibitem{Neves_Saa}Neves, J. C. S., Saa, A.: Regular rotating black holes and the weak energy condition. Phys. Lett. B \textbf{734}, 44 (2014) 

\bibitem{Lehners}Lehners, J. L.: Ekpyrotic and cyclic cosmology. Phys. Rept. \textbf{465}, 223 (2008) 

\bibitem{Brandenberger}Brandenberger, R. H.: The Matter bounce alternative to inflationary cosmology. arXiv:1206.4196

\bibitem{Steinhardt}Ijjas A., Steinhardt, P. J.: Inflationary paradigm in trouble after Planck2013. Phys. Lett. B \textbf{723}, 261 (2013)

\bibitem{Wu}Wu, K. K. S., Lahav, O., Rees, M. J.: The large-scale smoothness of the universe. Nature \textbf{397}, 225 (1999) 

\bibitem{Bolejko}Bolejko, K., Célérier,  M. N., Krasinski, A.: Inhomogeneous cosmological models: exact solutions and their applications. Class. Quantum Grav. \textbf{28}, 164002 (2011) 

\bibitem{Ruiz_Senovilla}Ruiz,  E., Senovilla, J. M. M.: General class of inhomogeneous perfect-fluid solutions. Phys. Rev. D \textbf{45}, 1995 (1992) 

\bibitem{Senovilla}Senovilla, J. M. M.: New class of inhomogeneous cosmological perfect-fluid solutions without big-bang singularity. Phys. Rev. Lett. \textbf{64}, 2219 (1990)

\bibitem{Nemiroff}Nemiroff, R. J. , Joshi, R., Patla, B. R.: An exposition on Friedmann cosmology with negative energy densities. JCAP \textbf{1506}, 006 (2015)  

\bibitem{Hernandez}Hernandez Jr, W. C.: Material sources for the Kerr metric. Phys. Rev. \textbf{159}, 1070 (1967)

\bibitem{Kim}Kim, C. W. , Song, J.: A proposed scale-dependent cosmology for the inhomogeneous cosmology. Int. J. Mod. Phys. D \textbf{5}, 293 (1996)

\bibitem{Wald}Wald, R.: General Relativity. The University of Chicago Press, Chicago (1984) 

\bibitem{Olmo}Olmo, G. J., Rubiera-Garcia, D., and Sanchez-Puente, A.: Geodesic completeness in a wormhole spacetime with horizons. Phys. Rev. D \textbf{92} 044047 (2015) 

\end{thebibliography}
\end{document}